\documentstyle[preprint,eqsecnum,aps]{revtex}

\begin{document}
\draft
\preprint{HEP/123-qed}
\title{Ground State Properties of the Two-Dimensional $t$-$J$ Model
}
\author{M.  Kohno}
\address{
Institute for Solid State Physics,\\
University of Tokyo, Roppongi, Minato-ku, Tokyo 106
}
\date{\today}
\maketitle
\begin{abstract}
The two-dimensional $t$-$J$ model in the ground state is investigated
by the power Lanczos method. The pairing-pairing correlation function
for $d_{x^2-y^2}$-wave symmetry is enhanced in the realistic parameter
regime for high-$T_{\rm c}$ superconductors. The charge susceptibility
$\chi_{\rm c}$ shows divergent behavior as $\chi_{\rm
c}\propto\delta^{-1}$ near half-filling for the doping concentration
$\delta$, indicating that the value of the dynamical exponent $z$ is
four under the assumption of hyperscaling. The peak height of the spin
structure factor $S_{\rm max}(Q)$ also behaves as $S_{\rm
max}(Q)\propto\delta^{-1}$ near half-filling, which leads to the
divergence of the antiferromagnetic correlation length $\xi_{\rm m}$
as $\xi_{\rm m}\propto\delta^{-1/2}$.  The boundary of phase
separation is estimated on the basis of the Maxwell
construction. Numerical results are compared with experimental
features observed in high-$T_{\rm c}$ cuprates.
\end{abstract}

\narrowtext

\section{Introduction}
\label{Introduction}

Since the discovery of high-$T_{\mbox{c}}$ cuprate
superconductors\cite{disc_HTC}, many microscopic models have been
proposed in order to explain the pairing mechanism. The best way to
understand the essential feature of the pairing mechanism would be to
find the simplest and realistic model that describes the low-energy
properties of the copper-oxide planes. The two-dimensional $t$-$J$
model is one of the candidates for the effective model for the
copper-oxide planes\cite{Anderson,Zhang}. As far as superconductivity
is concerned, E. Dagotto and J. Riera have obtained indications of
superconductivity by an unbiased diagonalization approach in the
region of $J/t \buildrel < \over \sim 3$ near
quarter-filling\cite{DagRiera}. More investigation is necessary in the
realistic region of couplings and densities in order to confirm that
the two-dimensional $t$-$J$ model can really be an effective model for
high-$T_{\rm c}$ superconductors. One of the purposes of this paper is
to investigate the relevancy of the two-dimensional $t$-$J$ model as a
low-temperature effective model for high-$T_{\rm c}$ cuprates.
\par
Also, strong electron correlation in low-dimensional systems is one of
the central issues in condensed matter physics. The Mott transition is
one of the remarkable consequences of strong correlation. In general,
there are two types of the Mott transitions for electron
systems\cite{Imada_Mott}. One is characterized by the vanishment of
the carrier density, the other is characterized by the divergence of
the carrier mass. A typical example which shows the first type of
transition is the free-fermion model on a lattice. In this case, the
dynamical exponent $z$ is two\cite{Imada_exp}. A typical example which
shows the second type of the Mott transition is the two-dimensional
Hubbard model\cite{Furukawa,Assaad_Hall}. It has been shown
numerically that the value of the dynamical exponent $z$ is four in
the case of the two-dimensional Hubbard
model\cite{Imada_exp,Assaad_loc}. A naive expectation is that the
two-dimensional $t$-$J$ model shows the same type of the Mott
transition as that of the two-dimensional Hubbard model, because the
$t$-$J$ model can be derived as an effective model for the Hubbard
model in the limit of $U\rightarrow\infty$. However, in the large
$J/t$ regime, the $t$-$J$ model shows different properties from those
of the Hubbard model. For example, if $J/t$ is so large that phase
separation occurs, the transition to an insulator is a first order
transition. In the region of $J/t$ slightly smaller than the
phase-separation boundary, it is expected that electrons (or holes)
form bound states due to the effective attractive forces that lead to
phase separation. It is interesting to investigate the critical
phenomena toward half-filling in this parameter regime. Hence, in this
paper, we investigate the Mott transition of the two-dimensional
$t$-$J$ model in the ground state near half-filling.
\par
Several obstacles to numerical calculations have prevented us from
getting the low-energy properties of the two-dimensional $t$-$J$
model. For example, the system size achieved by the exact
diagonalization is restricted to about 26 sites near half-filling. On
the other hand, Quantum Monte Carlo algorithms have a serious sign
problem. Recent progress of a Green's function Monte Carlo algorithm
(power Lanczos method\cite{PowerLanczos}) makes it possible for us to
investigate the ground state properties of the $t$-$J$ model in
relatively large systems. In this paper, we use the usual Lanczos
algorithm for clusters up to 20 sites and the power Lanczos method for
larger clusters.
\par
In Sec. \ref{Model}, the $t$-$J$ model is defined and the power
Lanczos method is briefly reviewed. In Sec. \ref{ResultDiscussion1},
we show the ground state energy as a function of filling and discuss
phase separation on the basis of the Maxwell construction. In
Sec. \ref{ResultDiscussion2}, numerical results on the pairing-pairing
correlation function are presented. In Sec. \ref{ResultDiscussion3},
the Mott transition at $J/t=0.5$ in the two-dimensional $t$-$J$ model
is discussed on the basis of the hyperscaling hypothesis.
Section \ref{Summary} is devoted to summary.

\section{Model and Method}
\label{Model}

The $t$-$J$ model is defined by the following Hamiltonian:
\begin{eqnarray}
   {\cal H}_{tJ} &=& {\cal H}_t+{\cal H}_J , \nonumber\\
   {\cal H}_t &=& -t\sum_{<i,j> \sigma}
(\tilde{c}^{\dagger}_{i\sigma}\tilde{c}_{j\sigma}+{\mbox{h.c.}}) , \nonumber\\
   {\cal H}_J &=& J\sum_{<i,j>}(\mbox{\boldmath $S_i\cdot S_j$}-\frac{1}{4}n_in_j),
\end{eqnarray}
where $\tilde{c}^{\dagger}_{i\sigma}$ denotes a creation operator of
an electron at site $i$ with spin $\sigma
(\sigma=\uparrow,\downarrow)$ with the constraint that no site is
doubly occupied, which is defined as
$\tilde{c}^{\dagger}_{i\sigma}\equiv (1-n_{i -
\sigma})c^{\dagger}_{i\sigma}$. The number operator $n_{i\sigma}$ is
defined as $n_{i\sigma}\equiv c^{\dagger}_{i\sigma}c_{i\sigma}$, using
the standard electron creation operator $c^{\dagger}_{i\sigma}$. The
spin operator at site $i$ is defined as $\mbox{\boldmath
$S$}_i\equiv\frac{1}{2}\sum_{\alpha\beta}c^{\dagger}_{i\alpha}\mbox{\boldmath
$\sigma_{\alpha\beta}$}c_{i\beta}$, where $\mbox{\boldmath
$\sigma_{\alpha\beta}$ }$ is the vector of Pauli matrices. The
summation ($\sum_{<i,j>}$) is taken over all nearest neighbor sites on
a square lattice.
\par
We adopt the power Lanczos method proposed by Y.C. Chen and
T.K. Lee\cite{PowerLanczos}. In the framework of this method, the
expectation value of an operator ${\cal O}$ in the ground state of a
Hamiltonian ${\cal H}$ is evaluated by the following equation:
\begin{equation}
   \langle{\cal O}\rangle = \lim_{p\rightarrow\infty}\langle 
p{\rm{L1}}|{\cal O}|p{\rm{L1}}\rangle/ 
\lim_{p'\rightarrow\infty}\langle p'{\rm{L1}}|p'{\rm{L1}}\rangle,
\end{equation}
where $|p{\rm{L1}}\rangle$ is the wavefunction defined as
$|p{\rm{L1}}\rangle\equiv{{\cal H}}^{p} |{\rm{L1}}\rangle$. 
The wavefunction $|{\rm{L1}}\rangle$ is defined as
$|{\rm{L1}}\rangle\equiv|{\rm trial}\rangle+c_1{{\cal H}}|{\rm trial}\rangle$,
where $|{\rm trial}\rangle$ is a trial wavefunction and $c_1$ is a
variational parameter.
if we set $c_1=0$, the power Lanczos method reduces to the power method.
\par
The variational wavefunction proposed by R. Valenti and
C. Gros\cite{ValentiGros} is employed as the trial wavefunction:
\begin{equation}
   |{\phi}\rangle = \prod_{ij}|r_i-r_j|^{\nu}{\rm{P_{G}}}
{\rm{P_{\it N}}}|{d\mbox{-wave}}\rangle,
\end{equation}
where $\nu$ is a variational parameter and $r_i$ represents the
real-space coordinate at site $i$. The Gutzwiller and $N$-particle
projection operators are denoted by ${\rm{P_{G}}}$ and ${\rm{P_{\it
N}}}$, respectively. The wavefunction $|d\mbox{-wave}\rangle$
represents the BCS-wavefunction in which the order parameter has
$d_{x^2-y^2}$-wave symmetry. It should be noted that the trial
wavefunction is in the subspace that both the total spin $S$ and the
total momentum $P$ are zero. Therefore the ground state properties
reported in this paper are within this subspace.
\par
One of the reasons why we use the above wavefunction $|{\phi}\rangle$ 
as a trial wavefunction 
is that this wavefunction gives the lowest energy 
among variational wavefunctions proposed so far, as far as we know, 
in the parameter regime we investigated. 
Another reason is that we can restrict the Hilbert space of simulation 
within the subspace of $S=0$ and $P=0$. This makes convergence faster.
The applicability of the power Lanczos method depends 
on the negative-sign ratio $r$ which is defined by $r\equiv (p-n)/(p+n)$, 
where $p$ and $n$ denote the number 
of positive and negative samples, respectively.
If the ratio $r$ is less than 0.1, it is difficult to obtain reliable results.
As a result, the applicability of the power Lanczos method is restricted 
to the power $p<p_{\rm r}$, 
where $p_{\rm r}$ denotes the power at which the ratio $r$ is about 0.1.
If the power required to reach convergence ($p_{\rm c}$) is 
larger than $p_{\rm r}$,
the power Lanczos method is not applicable.
If we use a wavefunction which has small overlap with the ground-state wavefunction,
it requires large power to reach the ground state.
We compare the speed of convergence of the power Lanczos method 
and the simple power method using the Gutzwiller wavefunction 
and $|{\phi}\rangle$ as a trial wavefunction. 
As shown in Fig.\ref{ene_converge}, 
the power Lanczos method requires smaller power $p$ 
than the simple power method,  
and the wavefunction $|{\phi}\rangle$ is superior 
to the Gutzwiller wavefunction as a trial wavefunction.
In this figure, the error in energy due to finite power $p$ is 
less than $0.2\%$ for $p>5$ 
by the power Lanczos method using $|{\phi}\rangle$,
although $p_{\rm c}$ becomes larger than $p_{\rm r}$
if we use the Gutzwiller wavefunction.
\par
We have checked convergent behavior in each simulation 
using $|{\phi}\rangle$ as a trial wavefunction.
The $p_{\rm c}$ becomes larger, if the system size becomes larger.   
For 50-site clusters, $p_{\rm c}$ is about eight for the energy to converge.
The $p_{\rm r}$ becomes smaller, if $J$ gets smaller.   
The most severe $p_{\rm r}$ in our simulation is about eight near half-filling 
for $J\simeq0.3$. 
As an example, we show the convergence of energy in a 50-site cluster 
with 42 electrons at $J=0.3$ in Fig.\ref{ene_converge42s50}.
We measure physical quantities at $p\simeq 8$, 
where we have checked in each simulation 
that the energy converges within a required accuracy.
As a check of convergence of physical quantities, 
we show the pairing-pairing correlation function
in a 20-site cluster with 18 electrons at $p=8$ in Fig.\ref{pair_convergence}.
\par
In the following sections, we show the numerical results in
finite-size clusters up to 104 sites. The boundary conditions are
chosen for the momentum configuration to be closed-shell.
We have typically run 1000-2000 Monte Carlo steps. Several hundred
branches are produced at each Monte Carlo step in the evaluation of
powers of ${\cal H}$.
\par
The filling $n$ is defined as $n\equiv N_{\rm e}/N_{\rm s}$, where
$N_{\rm e}$ is the number of electrons and $N_{\rm s}$ is the number
of sites. The doping concentration $\delta$ is defined as
$\delta\equiv 1-n$. The ground state energy per site at filling $n$ is
denoted by $e(n)$. Hereafter we set $t=1$ as the energy unit.
\narrowtext
\section{Phase Separation}
\label{ResultDiscussion1}
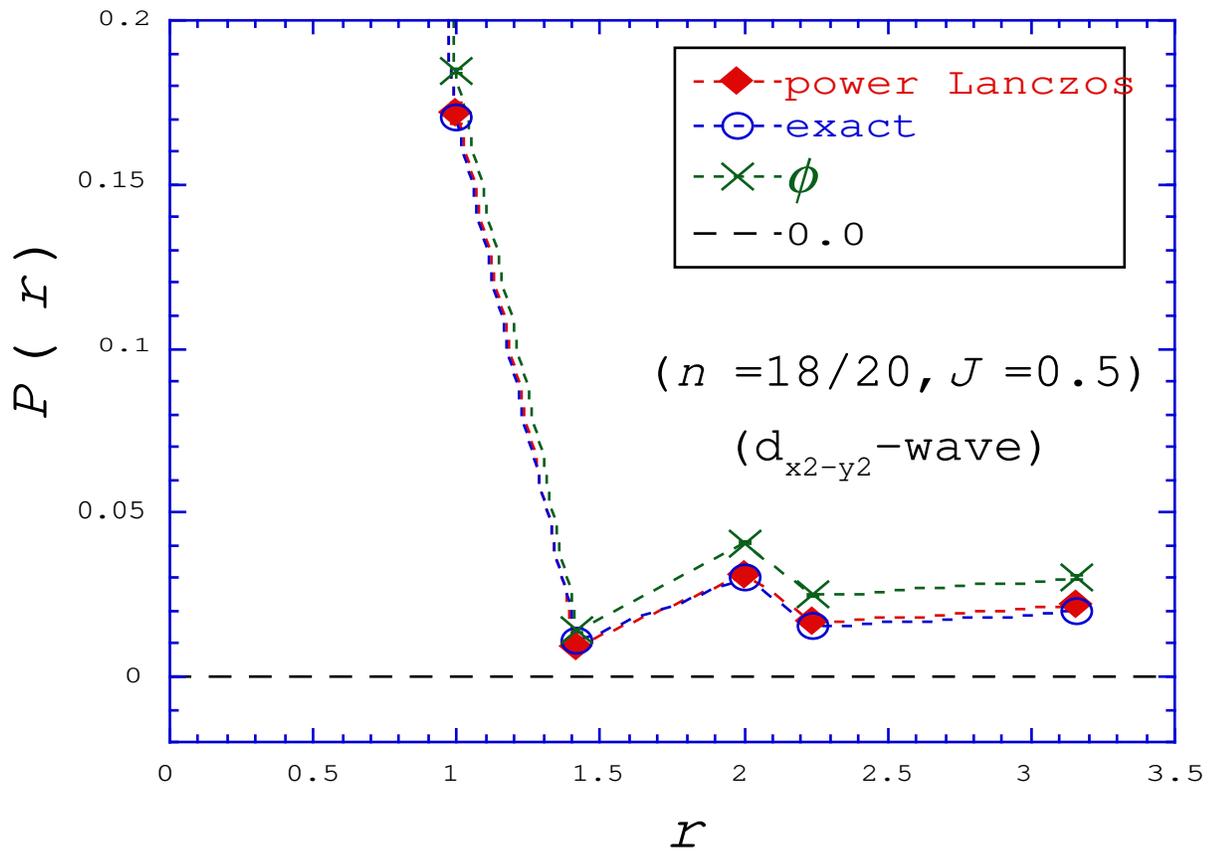
\begin{figure}
\caption{Ground state energy per site as a function of filling in the
two-dimensional free-fermion model on a square lattice. Dotted line
denotes a polynomial fit. Solid line denotes the ground state energy
per site in the thermodynamic limit.}
\label{ene_n_f}
\end{figure}

Before investigating phase separation of the two-dimensional $t$-$J$
model, we examine the finite-size effects on the ground state energy
in the free-fermion model on a square lattice, in which we can
calculate the exact ground state energy with any size of systems. We
calculate the ground state energy per site of the free-fermion model
in the same system sizes under the same boundary conditions as those
used in the $t$-$J$ model. We fit them as a function of filling by a
polynomial up to third order. As shown in Fig.\ref{ene_n_f}, the
fitting curve (dotted line) almost coincides with the curve in the
thermodynamic limit (solid line). The finite-size effects on the
ground state energy of the two-dimensional $t$-$J$ model is probably
not so different from those of the free-fermion model on a square
lattice. Actually, as shown in Fig.\ref{ene_n}, the data of the ground
state energy per site of the two-dimensional $t$-$J$ model in
finite-size clusters are well fitted by a polynomial as a function of
filling with small deviation from the fit, indicating that the
finite-size effects on the ground state energy are small.
\par
Figure \ref{ene_n} shows the ground state energy per site as a
function of filling in the two-dimensional $t$-$J$ model at $J=0.5,
1.0, 1.5, 2.0, 2.5$ and $3.0$. In this figure, the tangent from the
point at $n=1$ to the fitting curve gives a lower energy than the
fitting curve in the region of $n_{\rm c}<n<1$ as represented by the
solid line. Here $n_{\rm c}$ is the electron density at the point of
contact between the fitting curve and the tangent. Hence, we can
identify the region of phase separation as $n_{\rm c}<n<1$ on the
basis of the Maxwell construction\cite{ene_size_correction}. The
energy and the chemical potential of the phase-separated state are
given as the tangent and its slope, respectively. At $J=3.5$, we find
that the critical density $n_{\rm c}$ is zero. On the other hand, as
shown in Fig.\ref{ChemJ0.5}, the chemical potential at $J=0.5$ shows a
monotonically increasing behavior as a function of filling at least in
the region of $n\buildrel < \over \sim 0.95$, indicating that phase
separation does not occur at $J=0.5$. Hence, the phase-separation
boundary is obtained as in Fig.\ref{PD}. The critical $J$ above which
no stable homogeneous state exists is estimated as $J_{\rm
c_1}=3.4\pm0.1$. The critical $J$ below which no phase-separated state
exists is estimated as $J_{\rm c_2}=0.75\pm0.25$.
\par
The critical value of $J_{\rm c_1}$ is obtained more accurately in
Ref.\ref{HelMan} as $3.4367\pm 0.0001$ by solving the equation of
motion of two electrons. The numerical result in this paper is
consistent with it. The estimation of $J_{\rm c_2}$ is consistent with
that in Ref.\ref{Poil}($J_{\rm c_2}\simeq0.75$). In the intermediate
region of $n$, the phase-separation boundary estimated in this paper
is qualitatively similar to those in Ref.\ref{DagRiera} or
Ref.\ref{HTE_PS}, but quantitatively lower than them (Fig.\ref{PD}).
\par
In the following sections, we show the numerical results at $J=0.5$.
At this $J$, phase separation does not occur as discussed in this section.
\narrowtext
\section{Pairing-pairing correlation}
\label{ResultDiscussion2}
In this section, we show numerical results on the pairing-pairing
correlation functions $P_{\pm}(r)$ defined as
\begin{equation}
   P_{\pm}(r) \equiv \frac{1}{N_s}\sum_{r_0}
\langle\Delta_{\pm}(r_0)^{\dagger}\Delta_{\pm}(r_0+r)\rangle.
\end{equation}
Here, the singlet pairing operators $\Delta_{\pm}(r)$ are defined as
$\Delta(r)\equiv
c_{r\uparrow}(c_{r+\hat{x}\downarrow}+c_{r-\hat{x}\downarrow}\pm
c_{r+\hat{y}\downarrow}\pm c_{r-\hat{y}\downarrow})$, where $+$ and
$-$ correspond to extended $s$-wave and $d_{x^2-y^2}$-wave symmetry,
respectively and the unit vectors in $x$- and $y$- directions are
represented by $\hat{x}$ and $\hat{y}$, respectively.
\par In Fig. \ref{super50}(b), the pairing-pairing correlation
function with $d_{x^2-y^2}$-wave symmetry decays very little for
$n=0.84$, although the pairing-pairing correlation function with
$d_{x^2-y^2}$-wave symmetry quickly decays for $n=0.20$ as shown in
Fig. \ref{super50}(a).
\par
We define here the reduced pairing susceptibility as
\begin{equation}
\tilde{\chi}^{\rm P}_{\pm}\equiv\sum_{|r|> 2}P_{\pm}(r).
\end{equation}
 Figures \ref{super_n} (a) and (b) show the filling-dependence of
$\tilde{\chi}^{\rm P}_{\pm}/\tilde{N}_{\rm s}$, where $\tilde{N}_{\rm
s}$ is defined as $\tilde{N}_{\rm s}\equiv\sum_{|r|>2}1$. If the
superconducting long-range order exists, the value of
$\tilde{\chi}^{\rm P}_{\pm}/\tilde{N}_{\rm s}$ remains finite in the
thermodynamic limit. In Figs.\ref{super_n} (a) and (b), the
pairing-pairing correlation for $d_{x^2-y^2}$-wave symmetry is
enhanced in the region of $0.6\buildrel < \over \sim n\buildrel <
\over \sim 1$, and that for extended $s$-wave symmetry is a little
enhanced in the low-density regime.
\par
The numerical results showing that the $d_{x^2-y^2}$-wave component of
the pairing-pairing correlation is dominant near half-filling are
consistent with experimental indications, for example, the
measurements of the phase coherence in bimetallic YBCO-Pb dc
SQUIDs\cite{SQUID_experiment}.
\narrowtext
\section{Mott transition}
\label{ResultDiscussion3}
In Fig. \ref{ChemJ0.5}, the filling-dependence for the chemical
potential at $J=0.5$ is shown. The data of the chemical potential in
finite-size clusters are calculated as follows:
\begin{equation}
   \bar{\mu}(\bar{n}) \equiv \frac{e(n_1)-e(n_2)}{n_1-n_2},
\end{equation}
where $\bar{n}$ is taken as $\bar{n}=(n_1+n_2)/2$. Here, $n_1$ and
$n_2$ are taken to be adjacent closed-shell filling with boundary
conditions fixed. Here, boundary conditions at half-filling are
regarded as those under which the momentum configurations are
closed-shell in the free-fermion model on a square lattice. In the
thermodynamic limit, this definition of the chemical potential reduces
to the normal one: $\mu(n) \equiv \partial e(n)/\partial n$.
\par
We fit the data near half-filling in Fig.\ref{ChemJ0.5}(c)
as $\mu-\mu_{\rm c}\propto \delta^{\alpha}$ and
estimate $\mu_{\rm c}=1.31 \pm 0.03$ and $\alpha = 1.78 \pm 0.29$, 
which is close to $\alpha = 2$ reported 
in the case of the two-dimensional Hubbard model 
at $U=4$\cite{Furukawa}.
This suggests that the charge susceptibility $\chi_{\rm c}$ 
defined by $\chi_{\rm c}\equiv \partial n/\partial \mu$ 
diverges as $\chi_{\rm c}\propto
\delta^{-1}$ toward half-filling.
Actually the chemical potential is fitted well 
by the following form:
\begin{equation}
   |\mu-\mu_{\rm c}| \propto \delta^2,
\end{equation}
as denoted by the dashed line in Fig.\ref{ChemJ0.5}(a) and (b). 
In order to check this divergent behavior of the charge susceptibility,
we also investigate the doping dependence of chemical potential 
for $J=0.3$ and $J=0.4$. 
Figure \ref{ChemJ0.3} shows the same plot as in
Fig.\ref{ChemJ0.5} for $J=0.3$ and $J=0.4$.
From the fit of the numerical data in Fig.\ref{ChemJ0.3}(c), 
we estimate $\mu_{\rm c}=1.88 \pm 0.03$, $\alpha = 1.83 \pm 0.17$ for $J=0.3$ 
and $\mu_{\rm c}=1.57 \pm 0.02$, $\alpha = 1.98 \pm 0.07$ for $J=0.4$.
The numerical results for $J=0.3$ and $J=0.4$ also suggest 
that the charge susceptibility $\chi_{\rm c}$ diverges 
as $\chi_{\rm c}\propto \delta^{-1}$.
The divergent behavior of the charge
susceptibility is consistent with recent photo-emission
measurements\cite{photo2_experiment}.
\par
If the hyperscaling relations are satisfied, the charge susceptibility
$\chi_{\rm c}$ near the transition point to an insulator is written as
$\chi_{\rm c} \propto \delta^{-(z-d)/d}$, where $z$ is the dynamical
exponent and $d$ is the spatial
dimensionality\cite{Imada_exp}. Therefore the numerical results
suggest that the value of the dynamical exponent $z$ is four.
\par
We define the spin structure factor $S(k)$ as
\begin{equation}
   S(k) \equiv \frac{1}{3}\sum_{r}
\langle\mbox{\boldmath $S_0\cdot S_r$}\rangle {\rm e}^{{\rm i}kr}.
\end{equation}
Figure \ref{Skpeak}(a) shows the peak height of the spin structure factor
$S_{\rm max}(Q)$ as a function of filling.
The data near half-filling
can be fitted well by the following form:
\begin{equation}
   S_{\rm max}(Q) \propto \delta^{-1},
\end{equation}
as denoted by the dashed line(Fig.\ref{Skpeak}(a), (b) and (c)).
We fit the data near half-filling 
as $S_{\rm max}(Q)^{-1}\propto \delta^{\beta}$ and
estimate $\beta = 1.02 \pm 0.02$.
This suggests that the
antiferromagnetic correlation length $\xi_{\rm m}$ diverges toward
half-filling as
\begin{equation}
   \xi_{\rm m} \propto \delta^{-1/2},
\end{equation}
under the assumption that the spin-spin correlation behaves as 
$\langle\mbox{\boldmath $S_0\cdot S_r$}\rangle\propto 
{\rm e}^{{\rm i}Qr}\cdot {\rm e}^{-r/\xi_{\rm m}}$\cite{Furukawa}. 
This behavior of the correlation length has been reported on the 
two-dimensional Hubbard model at $U=4$\cite{Furukawa} and 
is consistent with the observation by neutron scattering 
experiments\cite{Sk_experiment}.
\par
Under the assumption of the existence of the single
characteristic length scale $\xi$ that is related to critical
phenomena, the hyperscaling theory has predicted that the length scale
$\xi$ diverges as $\xi \propto \delta^{-1/d}$ toward the critical
point, where $d$ is the spatial dimensionality\cite{Imada_exp}. The
numerical results shown above support the scaling hypothesis and
suggest that the Mott transition in the two-dimensional $t$-$J$ model
at $J=0.5$ is characterized by the dynamical exponent $z=4$, which is
the same as in the case of the two-dimensional Hubbard model at
$U=4$\cite{Imada_exp,Assaad_loc}.
\par
\narrowtext
\section{Summary}
\label{Summary}
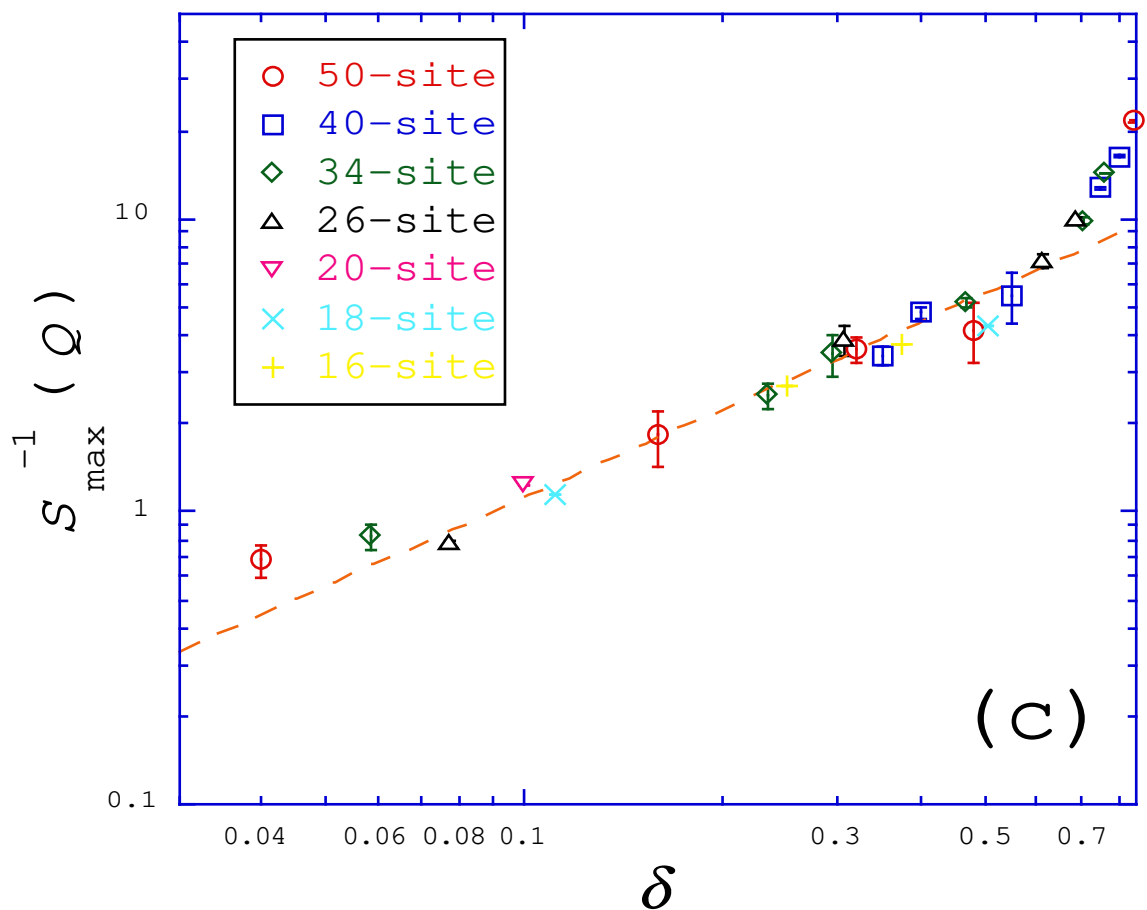
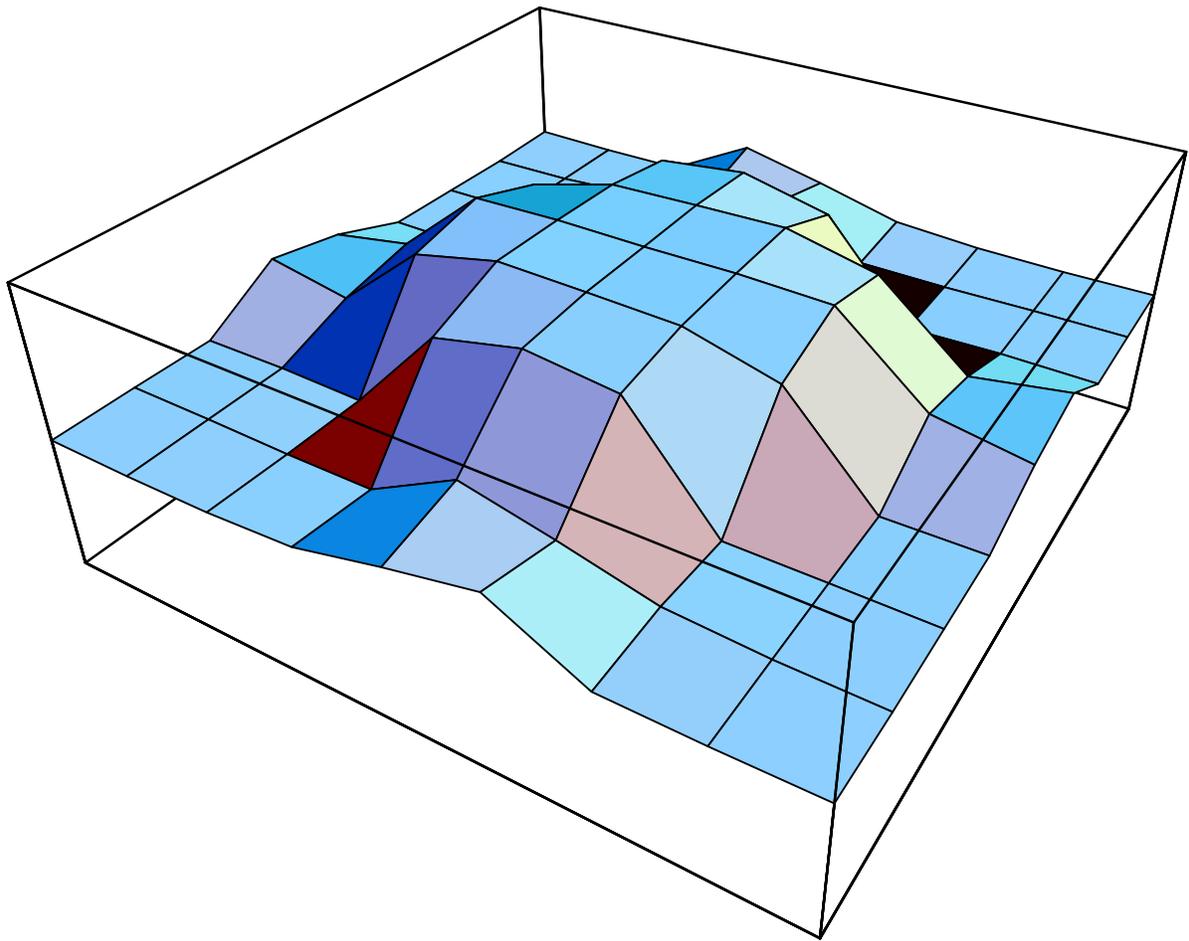
\begin{figure}
\caption{Momentum distribution function for $n=0.84$ at $J=0.5$ in a
50-site cluster. In the inset, solid and dashed lines denote the Fermi
Surface and the Brillouin zone boundary, respectively. The center is
the $\Gamma$-point.}
\label{nk}
\end{figure}

Numerical results presented in this paper are consistent with the
following experimental features found in the high-$T_{\rm c}$ oxides:
(i) the $d_{x^2-y^2}$-wave symmetry of the superconducting order
parameter in the region of moderate doping, which is suggested by the
measurements of the phase coherence in bimetallic YBCO-Pb dc
SQUIDs\cite{SQUID_experiment} (Sec. \ref{ResultDiscussion2}), (ii) the
doping dependence of the antiferromagnetic correlation length near
half-filling ($\xi_{\rm m}\propto \delta^{-1/2}$) observed in neutron
scattering experiments\cite{Sk_experiment}
(Sec. \ref{ResultDiscussion3}), (iii) the large Fermi surface behavior
in the region of moderate doping\cite{photo_experiment} (Fig.\ref{nk})
and (iv) divergent behavior of the charge susceptibility suggested
by photo-emission experiments\cite{photo2_experiment}
(Sec. \ref{ResultDiscussion3}).
\par
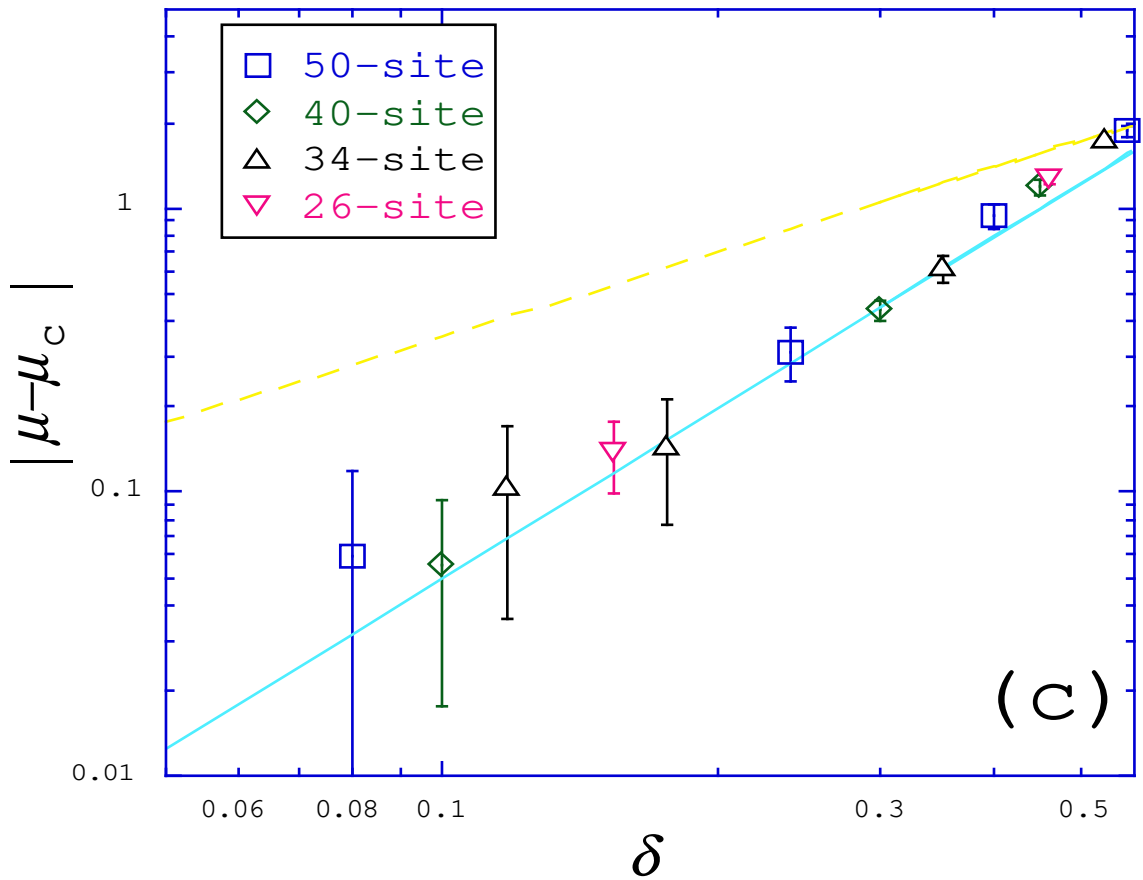
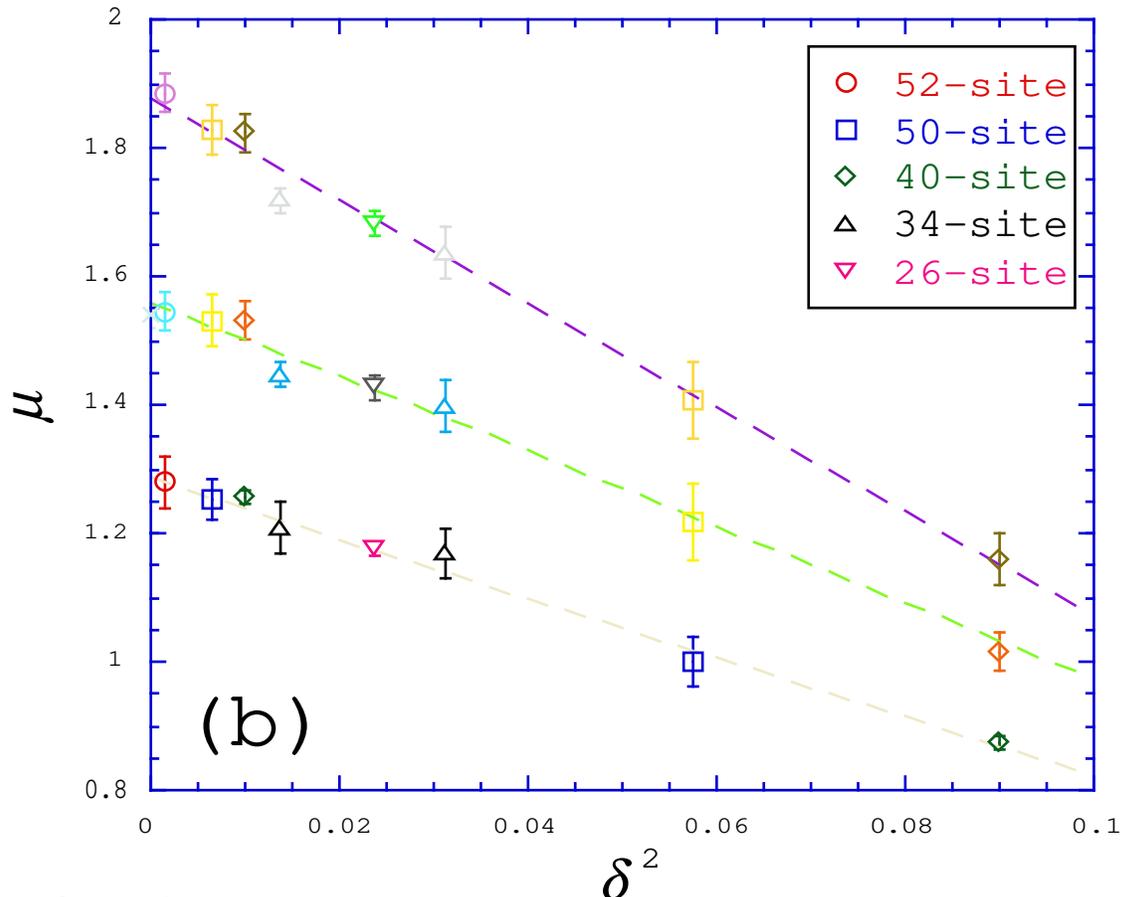
\begin{figure}
\caption{Chemical potential as a function of filling at $J=0.3$, $0.4$
and $0.5$ starting from above, (a) linear plot, (b) $\mu$ vs $\delta^2$ plot. 
(c)Log-log plot for $J=0.3$ (open symbols) and $J=0.4$ (solid symbols).
Dashed line in (a) and (b) denotes a fit as $|\mu-\mu_c|\propto\delta^2$.
The point denoted by a cross is obtained 
by a Green's function Monte Carlo in a 10$\times$10-site cluster 
with two holes taken from Ref.\protect\ref{J4GFMC2holes}.
Dashed and solid lines in (c) correspond 
to the cases of the dynamical exponent $z=2$ and $z=4$, respectively.}
\label{ChemJ0.3}
\end{figure}
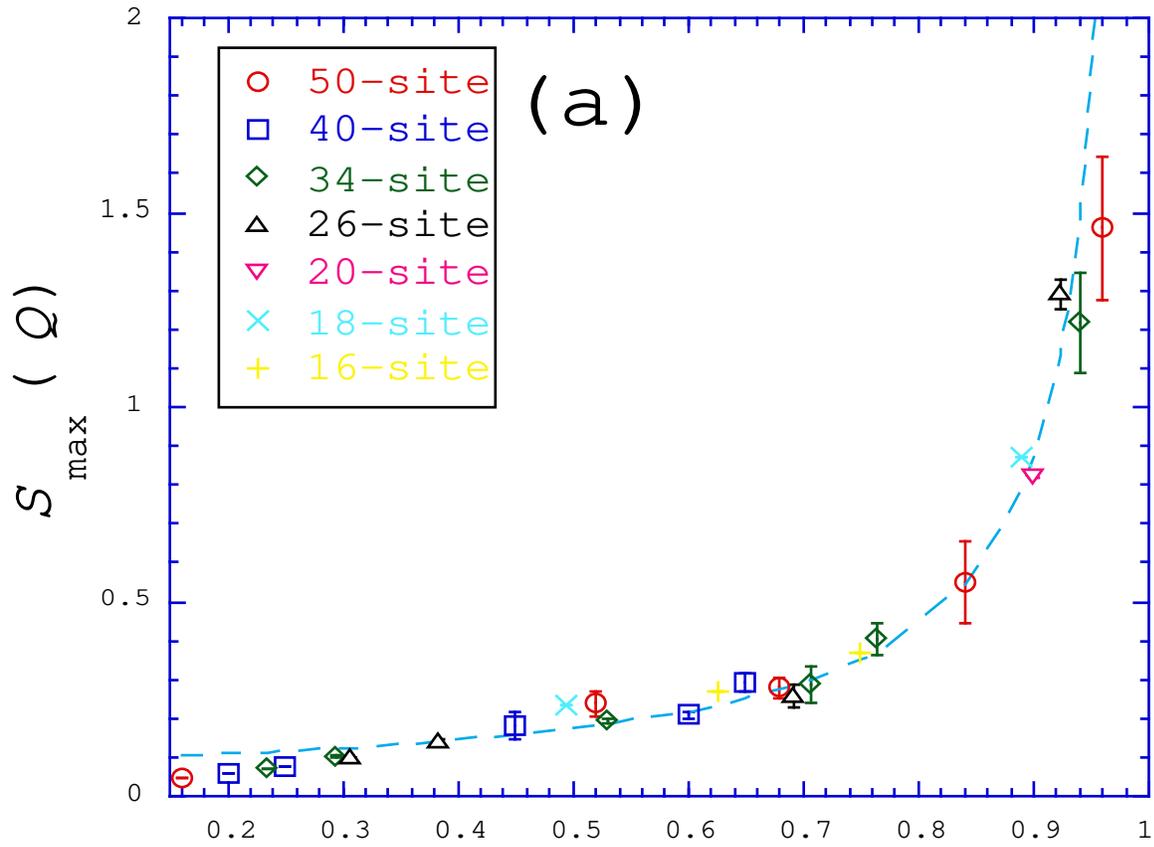
\begin{figure}
\caption{Peak height of the spin structure factor $S_{\rm max}(Q)$ as
a function of filling at $J=0.5$, (a) linear plot, 
(b) $S_{\rm max}(Q)^{-1}$ vs $\delta$ plot and (c) log-log plot.
Dashed line in (a), (b) and (c)  denotes a fit in the form of $S_{\rm max}(Q) \propto \delta^{-1}$.}
\label{Skpeak}
\end{figure}
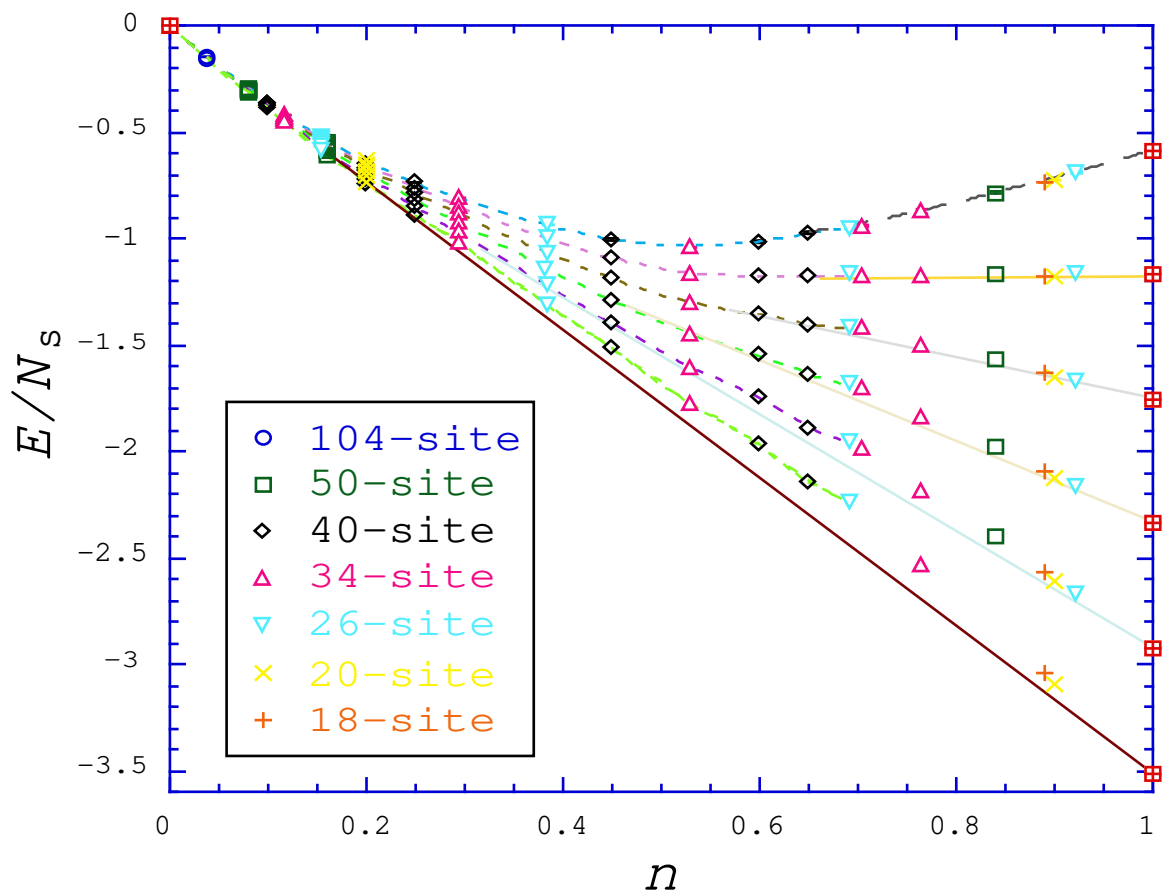
\begin{figure}
\caption{Schematic phase diagram of the two-dimensional $t$-$J$ model
in the ground state. Solid diamonds denote the phase-separation
boundary obtained in Sec. \protect\ref{ResultDiscussion1}. In the
higher-density region than the open diamond, the pairing-paring
correlation for $d_{x^2-y^2}$-wave symmetry is enhanced. It should be
noted that a ferromagnetic phase may exist in the small $J$ region
($J\buildrel < \over \sim 0.1$) suggested in Refs.\protect\ref{HTE_PS}
and \protect\ref{HTE_Ferro}. This ferromagnetic phase is beyond the
scope of this paper.}
\label{PD}
\end{figure}

In summary, numerical results on the two-dimensional $t$-$J$ model
have been reported. The boundary of phase separation is estimated on
the basis of the Maxwell construction (Fig.\ref{PD}). The
pairing-pairing correlation for $d_{x^2-y^2}$-wave symmetry is
enhanced in the region of $0.6 \buildrel < \over \sim n\buildrel <
\over \sim 1$ at $J=0.5$ (Fig.\ref{super_n}). The charge
susceptibility $\chi_{\rm c}$ shows divergent behavior as $\chi_{\rm
c}\propto\delta^{-1}$ toward half-filling, indicating that the value
of the dynamical exponent $z$ is four (Figs.\ref{ChemJ0.5} and
\ref{ChemJ0.3}). The peak height of the spin structure factor $S_{\rm
max}(Q)$ diverges toward half-filling as $S_{\rm max}(Q)\propto
\delta^{-1}$ (Fig.\ref{Skpeak}). This leads to the divergence of the
antiferromagnetic correlation length as $\xi_{m}\propto
\delta^{-1/2}$.

\narrowtext
\acknowledgments
The author would like to thank F.F. Assaad, F.V. Kusmartsev, M. Imada,
M. Takahashi, N. Furukawa and K. Kusakabe for helpful discussions and
useful comments. The author also thanks T. Kawarabayashi and
E. Williams for reading of the manuscript. The author would like to thank 
A. Ino for sending me the experimental data for the charge susceptibility. 
The exact diagonalization 
program is partly based on the subroutine package "TITPACK Ver.2"
coded by H. Nishimori and partly on the subroutines coded by
K. Kusakabe. Part of the exact diagonalization calculations were
performed on the Fujitsu VPP500 of the Supercomputer Center of the
Institute for Solid State Physics, Univ. of Tokyo. Part of the Monte
Carlo calculations were carried out on the Intel Japan PARAGON at the
Institute for Solid State Physics, Univ. of Tokyo.

\begin{figure} 
\caption{Energy per site as a function of power $p$ 
for the 20-site system with 18 electrons at $J=0.5$. 
The percentage of the error from the ground state energy corresponds 
to the vertical scale on the right. 
Open circles and open diamonds denote the data obtained 
by the power method and the power Lanczos method, respectively, 
using the Gutzwiller wavefunction as a trial wavefunction. 
Solid symbols denote the data by using
$|{\phi}\rangle$ as a trial wavefunction.}
\label{ene_converge}
 \end{figure}

\begin{figure} 
\caption{The same plot as in Fig.\protect\ref{ene_converge} but 
for the 50-site system with 42 electrons at $J=0.3$. 
Inset shows the negative-sign ratio $r$ defined in the text. 
The dashed line represents $r=0.1$.}
\label{ene_converge42s50}
\end{figure}

\begin{figure}
\caption{Pairing-pairing correlation function for $d_{x^2-y^2}$-wave
symmetry in a 20-site cluster with 18 electrons at $J=0.5$. Crosses
denote the data by $|{\phi}\rangle$ without the power method. Solid
diamonds denote the data by the power Lanczos method at $p=8$, using
$|{\phi}\rangle$ as a trial wavefunction. Open circles denote the
exact data obtained by the exact diagonalization.}
\label{pair_convergence}
\end{figure}

\begin{figure}
\caption{Ground state energy per site as a function of filling in the
two-dimensional $t$-$J$ model at $J=0.5, 1.0, 1.5, 2.0, 2.5$ and
$3.0$, starting from above. Dotted line denotes the same fit as in
Fig.\protect\ref{ene_n_f}, using data points from $n=0$ to
$n\simeq0.7$. Solid line denotes the expected ground state energy per
site of the phase-separated state which is determined on the basis of
the Maxwell construction. Dashed line at $J=0.5$ is obtained by
integrating the fit (dashed line) in Fig.\protect\ref{ChemJ0.5}.}
\label{ene_n}
\end{figure}

\begin{figure}
\caption{Pairing-pairing correlation function as a function of
distance for (a) $n=0.20$ and (b) $n=0.84$ at $J=0.5$ in 50-site
clusters. Crosses and solid diamonds denote the pairing-pairing
correlation functions for extended $s$-wave symmetry and
$d_{x^2-y^2}$-wave symmetry, respectively.}
\label{super50}
\end{figure}

\begin{figure}
\caption{Reduced pairing susceptibility per site ($\tilde{\chi}^{\rm
P}_{\pm}/\tilde{N}_{\rm s}$) as a function of filling at $J=0.5$ for
(a) extended $s$-wave symmetry and (b) $d_{x^2-y^2}$-wave symmetry,
where $\tilde{\chi}^{\rm P}_{\pm}\equiv\sum_{|r|>2}P_{\pm}(r)$ and
$\tilde{N}_{\rm s}\equiv\sum_{|r|>2}1$.}
\label{super_n}
\end{figure}

\begin{figure}
\caption{Chemical potential as a function of filling at
$J=0.5$, (a) linear plot, (b) $\mu$ vs $\delta^2$ plot and 
(c) log-log plot. 
Dotted line in (a) and (b) is obtained by differentiating the fit 
at $J=0.5$ (dotted line) in Fig.\protect\ref{ene_n}. 
Dashed line in (a) and (b) denotes a fit as $|\mu-\mu_c|\propto\delta^2$.
Dashed and solid lines in (c) correspond 
to the cases of the dynamical exponent $z=2$ and $z=4$, respectively.}
\label{ChemJ0.5}
\end{figure}

\begin{references}
\bibitem{ene_size_correction} The finite-size effects on the energy in
the phase-separated state may be much larger than those in a
homogeneous state because of macroscopic fluctuations. The
system sizes examined in this paper may be insufficient to obtain an
accurate value for the energy of the phase-separated state. Therefore
we use the Maxwell construction to determine it.
\label{ene_size_correction}
\end{references}
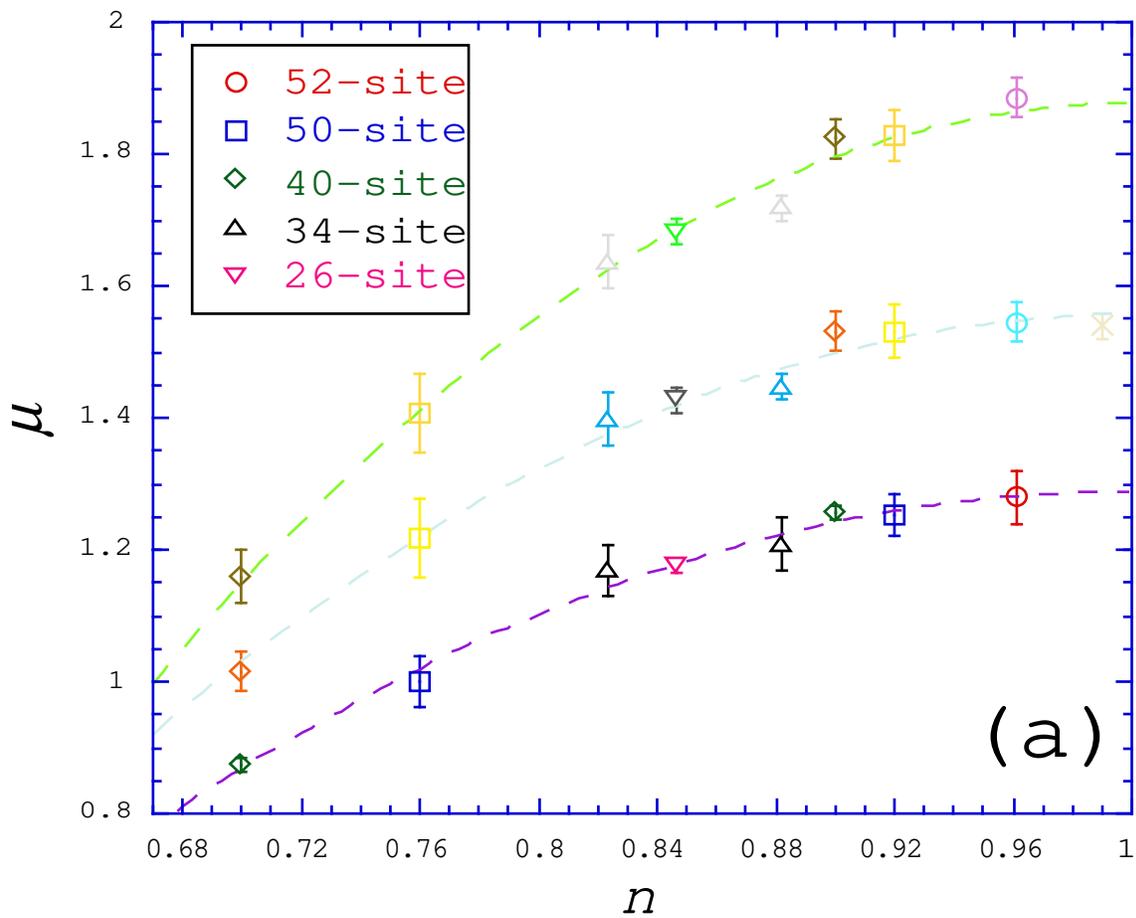
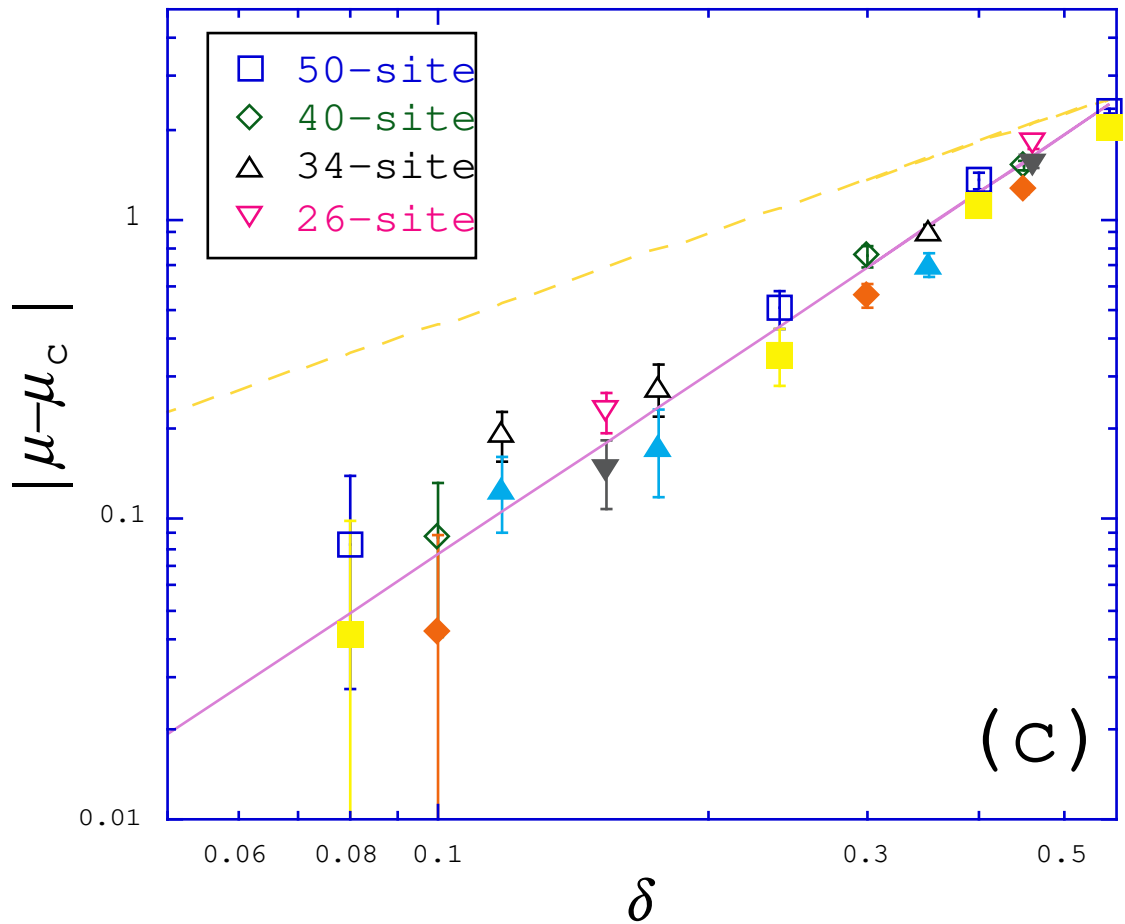
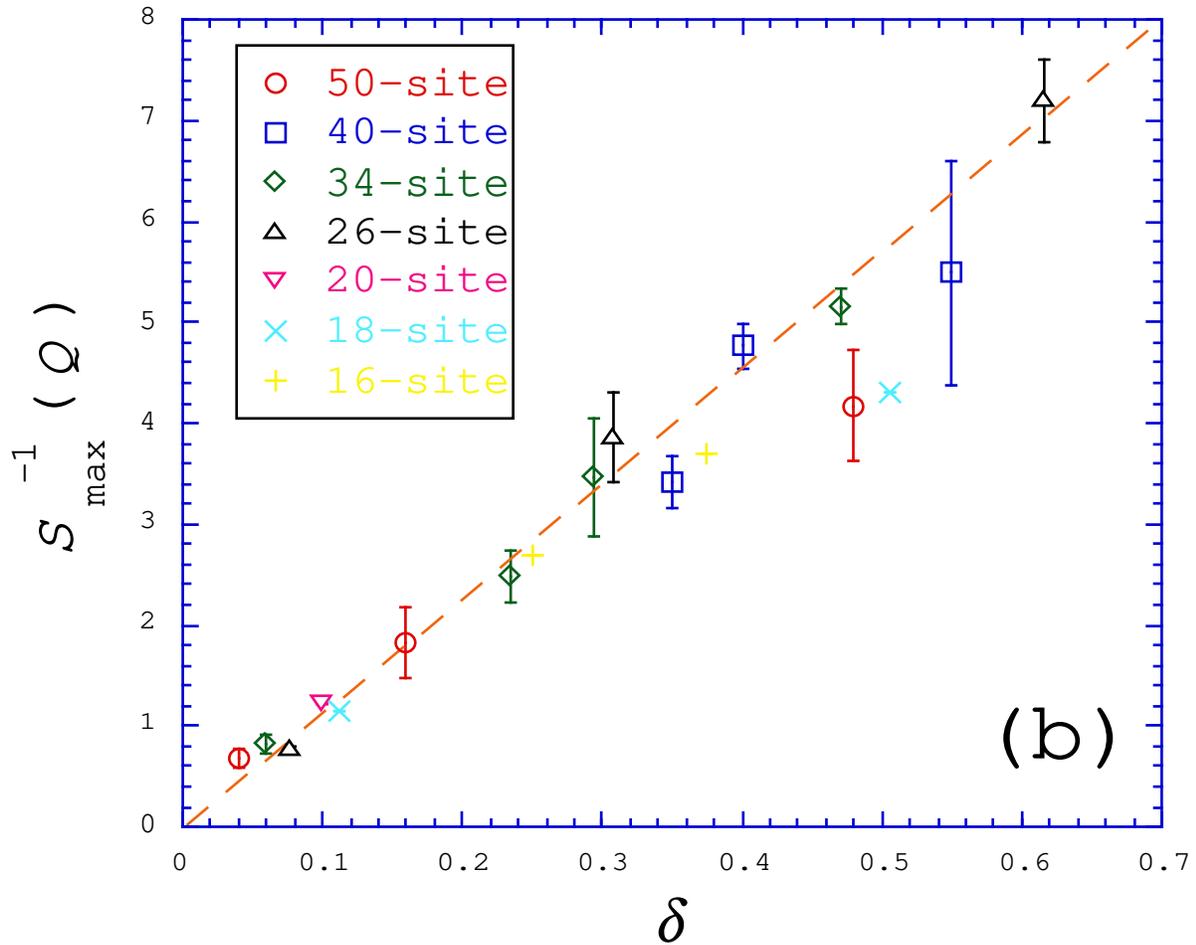
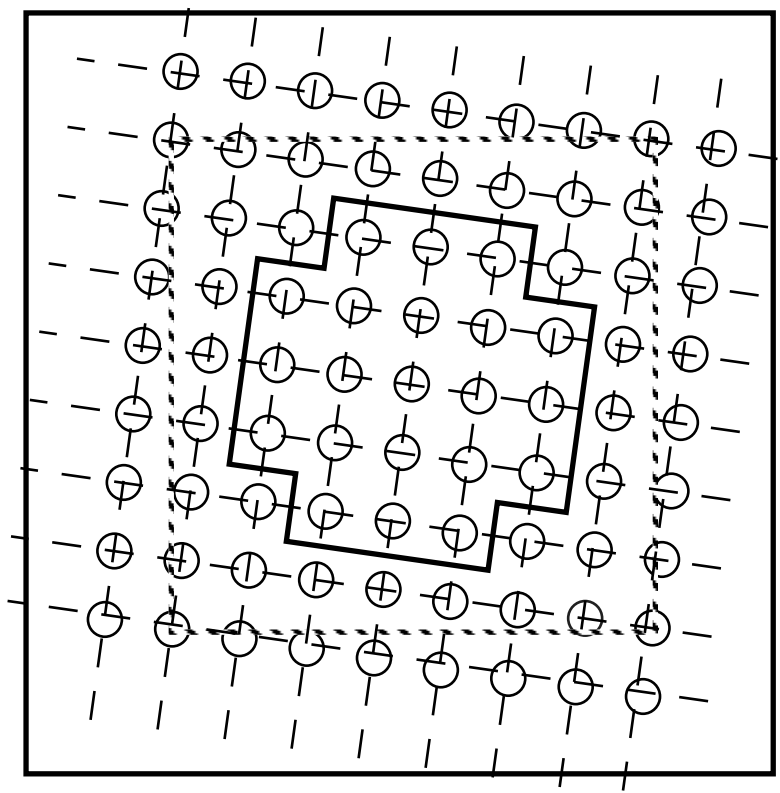
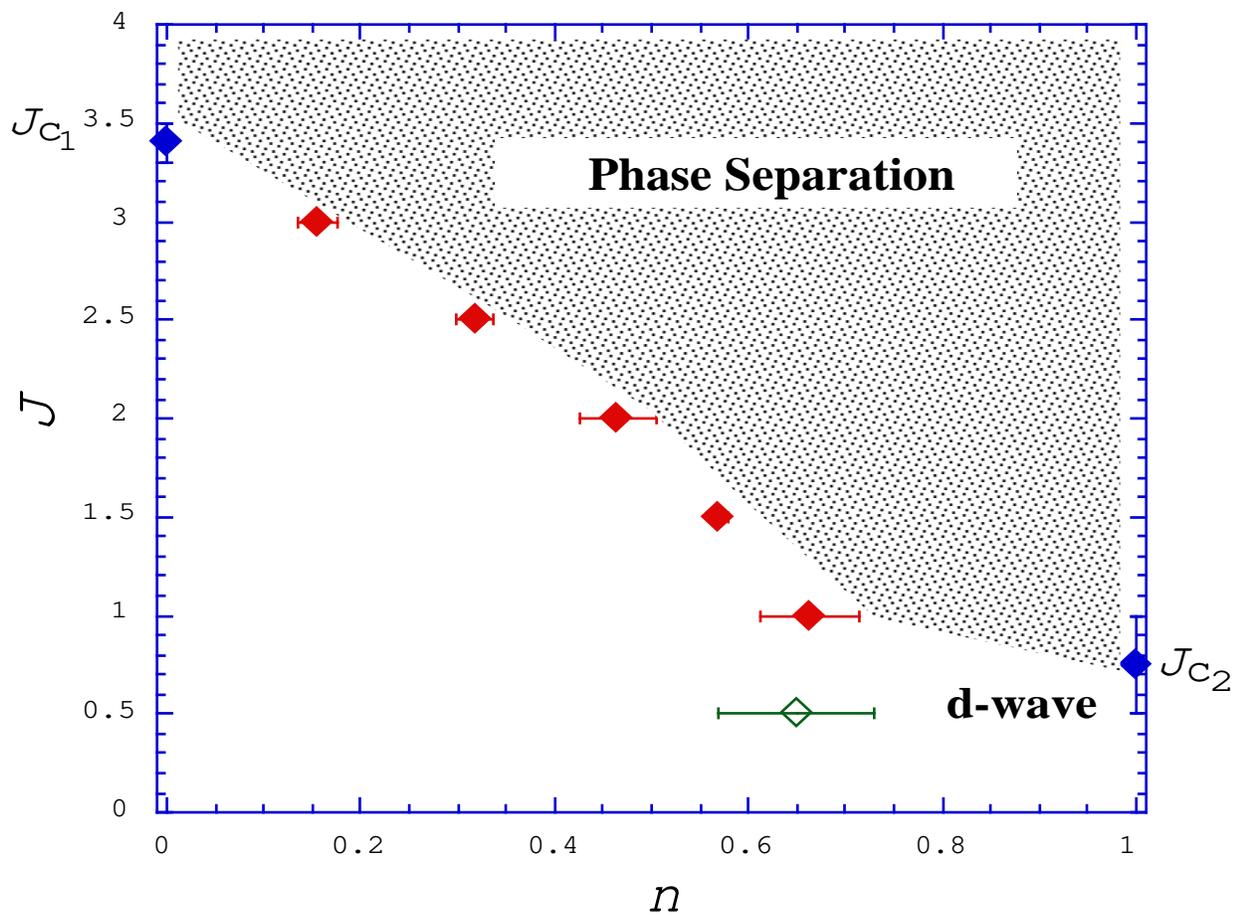
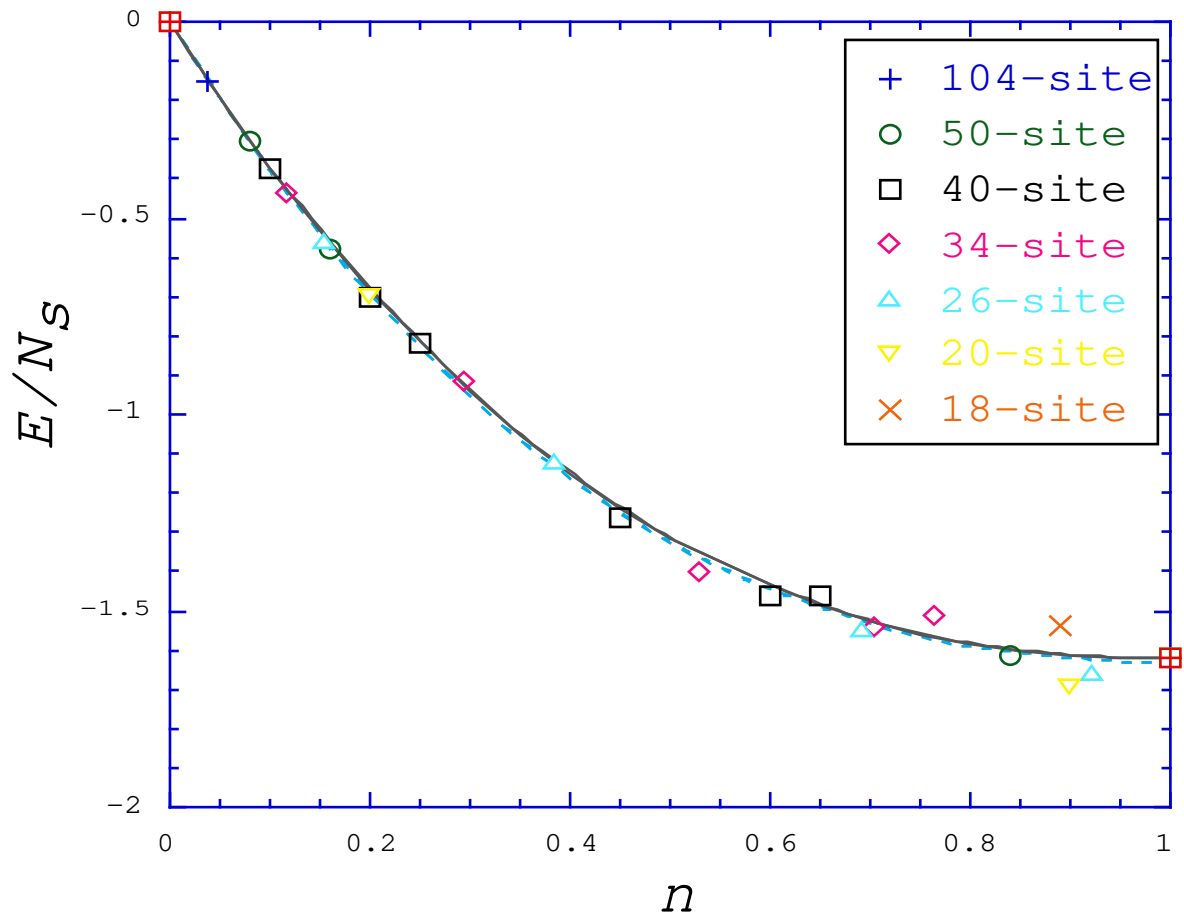
\end{document}